\documentclass[twocolumn]{aastex7}

\usepackage{amsmath}
\usepackage[T1]{fontenc}
\usepackage{multirow}
\usepackage{hyperref}
\usepackage{soul}

\newcommand\teff{$T_{\mathrm{eff}}$}

\begin{document}

\title{Scaling K2 VIII: Short-Period Sub-Neptune Occurrence Rates Peak Around Early-Type M Dwarfs}

\author[0000-0003-3702-0382]{Kevin K.\ Hardegree-Ullman}
\affiliation{Caltech/IPAC-NASA Exoplanet Science Institute, 1200 E.\ California Blvd., MC 100-22, Pasadena, CA 91125, USA}
\email[show]{kevinkhu@caltech.edu}

\author[0000-0003-4500-8850]{Galen J.\ Bergsten}
\affil{Lunar and Planetary Laboratory, The University of Arizona, Tucson, AZ 85721, USA}
\email[]{gbergsten@arizona.edu}

\author[0000-0002-8035-4778]{Jessie L.\ Christiansen}
\affiliation{Caltech/IPAC-NASA Exoplanet Science Institute, 1200 E.\ California Blvd., MC 100-22, Pasadena, CA 91125, USA}
\email[]{christia@ipac.caltech.edu}

\author[0000-0003-1848-2063]{Jon K.\ Zink}\altaffiliation{NASA Sagan Fellow}
\affiliation{Department of Astronomy, Caltech, 1200 E.\ California Blvd, Pasadena, CA 91125}
\email[]{jzink@caltech.edu}

\author[0000-0002-6673-8206]{Sakhee Bhure}
\affil{Centre for Astrophysics, University of Southern Queensland, Toowoomba, QLD 4350, Australia}
\email[]{sakhee.bhure@unisq.edu.au}

\author[0000-0001-8153-639X]{Kiersten M. Boley}
\altaffiliation{NASA Sagan Fellow}
\affil{Earth \& Planets Laboratory, Carnegie Institution for Science, Washington, DC, 20015, USA}
\email{kboley@carnegiescience.edu}

\author[0000-0002-3853-7327]{Rachel B. Fernandes}
\altaffiliation{President’s Postdoctoral Fellow}
\affil{Department of Astronomy \& Astrophysics, 525 Davey Laboratory, The Pennsylvania State University, University Park, PA 16802, USA}
\affil{Center for Exoplanets and Habitable Worlds, 525 Davey Laboratory, The Pennsylvania State University, University Park, PA 16802, USA}
\email{rbf5378@psu.edu}

\author[0000-0002-8965-3969]{Steven Giacalone}
\altaffiliation{NSF Astronomy and Astrophysics Postdoctoral Fellow}
\affiliation{Department of Astronomy, California Institute of Technology, Pasadena, CA 91125, USA}
\email{giacalone@astro.caltech.edu}

\author[0000-0002-1480-9041]{Preethi R.\ Karpoor}
\affil{Department of Astronomy and Astrophysics, UC San Diego, La Jolla, CA 92092, USA}
\email[]{pkarpoor@ucsd.edu}

\begin{abstract}

We uniformly combined data from the NASA Kepler and K2 missions to compute planet occurrence rates across the entire FGK and M dwarf stellar range. The K2 mission, driven by targets selected by guest observers, monitored nine times more M dwarfs than the Kepler mission. Combined, Kepler and K2 observed 130 short-period ($P=1-40$ days) Earth to Neptune-sized candidate planets orbiting M dwarfs. K2 observed 3.5 times more of these planets than Kepler for host stars below 3700~K. Our planet occurrence rates show that short-period sub-Neptunes peak at $3750^{+153}_{-97}$~K and drop for cooler M dwarfs. A peak near this location was predicted by pebble accretion planet formation models and confirmed here by observations for the first time. Super-Earths continue to increase in occurrence toward cooler stars and show no clear evidence of a peak in the host star range considered here (3200~K--6900~K). Our observations provide critical input to further refine planet formation models. We strongly recommend further study of mid-to-late M dwarfs with TESS and soon the Nancy Grace Roman Space Telescope and PLATO to identify additional small planet trends.
\end{abstract}


\section{Introduction} \label{sec:intro}

The primary goal of the Kepler mission was to determine how common Earth-sized planets are in the habitable zones of Sun-like FGK stars \citep{Borucki2010}. After four years of staring at the same patch of sky, two of the four reaction wheels used to keep the pointing stable on Kepler had failed, and the K2 mission was born \citep{Howell2014}. K2 conducted another four-year survey, successfully observing 18 separate fields along the ecliptic plane for $\sim$80 days each, driven by community-provided target lists.

While numerous studies attempted to achieve the original Kepler mission goal, few reliable habitable zone Earth-sized planets were found, making planet occurrence rates for this population model-dependent \citep{Bergsten2022}. Nonetheless, Kepler was transformative in exoplanet demographics, giving us crucial information about the close-in exoplanet population\footnote{See \url{https://exoplanetarchive.ipac.caltech.edu/docs/occurrence_rate_papers.html} for a growing list of exoplanet occurrence rate papers.}. The Scaling K2 project was created to push exoplanet demographics beyond the Kepler field, probing deeper into the short-period planet population across different regions of the galaxy and in different stellar populations.

Thus far, in Scaling K2 papers I--VII, we have compiled a homogeneous stellar catalog and identified and validated the planet radius valley beyond the Kepler field for the first time \citep[Scaling K2 I,][]{Hardegree-Ullman2020}, developed an automated planet detection pipeline with completeness and reliability metrics \citep[Scaling K2 II,][]{Zink2020a} and tested these tools on K2 Campaign 5 \citep[Scaling K2 III,][]{Zink2020b}, generated a uniform exoplanet candidate catalog for Campaigns 1--8 and 10--18 \citep[Scaling K2 IV,][]{Zink2021}, validated 60 new planets from the candidate catalog \citep[Scaling K2 V,][]{Christiansen2022}, computed FGK planet occurrence rates for K2 and compared them to the short-period Kepler planet population, identifying a trend of fewer small, short-period planets at high galactic latitude \citep[Scaling K2 VI,][]{Zink2023}, and computed a high hot sub-Neptune occurrence rate for young planets in the Praesepe and Hyades clusters, favoring a core-powered mass-loss scenario sculpting the planet radius valley \citep[Scaling K2 VII,][]{Christiansen2023}.

In \citet{Zink2023}, analysis of the Kepler and K2 FGK sample was conducted by combining information from the two surveys in a mostly homogeneous nature, relying on the K2 stellar and planet samples from \citet{Hardegree-Ullman2020} and \citet{Zink2021}, and Kepler data from \citet{Berger2020a} and \citet{Berger2020b}. While the planet samples from Kepler and K2 were derived in largely the same manner, the stellar samples were derived using two different methods. Notably absent from \citet{Zink2023} was a calculation of planet occurrence rates for M dwarfs. \citet{Dressing2015} and \citet{Mulders2015a,Mulders2015b} used Kepler data to identify that small planets were much more abundant around early-type M dwarfs than FGK stars, but to-date, no comprehensive analysis has been done with K2 M dwarfs. K2 observed about \textit{nine} times more M dwarfs than Kepler, making this a potentially rich, under-explored area of parameter space.

In this paper, we explore exoplanet demographics for the combined Kepler and K2 dataset for all FGK and M stars, including a homogeneous treatment of stellar properties. In Section~\ref{sec:sample} we describe our methodology for placing Kepler and K2 stars and exoplanets on the same scale to enable combined exoplanet demographics calculations. In Section~\ref{sec:planets} we describe our planet sample, followed by our planet occurrence rate calculations in Section~\ref{sec:rates}. We compare our new occurrence rates to planet formation models and discuss future directions of combined transit occurrence rate studies in Section~\ref{sec:discussion}, and finally conclude in Section~\ref{sec:conclusions}.

\section{Stellar Sample} \label{sec:sample}
Combining data from different surveys necessitates uniform analysis across datasets. The first step in this process is homogeneous stellar classification. In order to place Kepler and K2 targets on the same scale, we used photometry from Gaia DR3 \citep{GaiaCollaboration2023} and 2MASS \citep{Cutri2003,Skrutskie2006}, and probabilistic distance measurements based on Gaia DR3 from \citet{Bailer-Jones2021}. We started with the full Kepler and K2 target tables\footnote{Kepler stellar and K2 target tables from the Exoplanet Archive \citep{Christiansen2025}: \url{https://exoplanetarchive.ipac.caltech.edu/cgi-bin/TblSearch/nph-tblSearchInit?app=ExoTbls&config=keplerstellar} and \url{https://exoplanetarchive.ipac.caltech.edu/cgi-bin/TblSearch/nph-tblSearchInit?app=ExoTbls&config=k2targets}. Accessed 10 July 2024.}, which yielded 200,038 unique Kepler and 372,289 unique K2 targets. Next, we used the CDS cross-match service \citep{Boch2012,Pineau2020} to perform a 1\arcsec\ positional cross-match between Gaia DR3 and Kepler and K2 targets. The CDS cross-match service propagates the Gaia DR3 J2016 epoch positions and proper motions to J2000 for the cross-match. To further minimize false matches, we only kept targets with an absolute magnitude difference of $<1$ between the Gaia $G$-band and Kepler $Kp$-band. This left 195,045 unique Kepler and 257,850 unique K2 targets. We dropped targets without complete Gaia or 2MASS photometry or distances from \citet{Bailer-Jones2021}, and only kept K2 targets with a \texttt{k2\_type} flag of STAR. This left us with 189,595 unique Kepler and 235,880 unique K2 targets.

\subsection{Stellar Parameters}
The most fundamental stellar parameters we derived for our uniform catalogs are stellar effective temperature \teff, surface gravity $\log g$, metallicity [Fe/H], luminosity $L_{\star}$, radius $R_{\star}$, and mass $M_{\star}$. We largely adopted the methodology outlined in \citet{Hardegree-Ullman2020} (Sections 3 and 4) to derive stellar parameters, which we describe below. The main update we made in the following analysis was to use a color-temperature relationship to derive \teff, instead of using a spectroscopic training set. The color-temperature relationship is broader in its coverage of M dwarf temperatures than the spectroscopic training set used in \citet{Hardegree-Ullman2020}.

Following the methodology outlined in \citet{Hardegree-Ullman2023} (Section 2.4), we fit a 4th-order polynomial\footnote{The polynomial order was selected from the lowest Bayesian Information Criterion after iterating over polynomial order fits between 2 and 15.} to the Gaia $G_{BP}-G_{RP}$ versus stellar effective temperature (\teff) from the updated Table 5 of \citet{Pecaut2013}\footnote{Table maintained at \url{https://github.com/emamajek/SpectralType/blob/master/EEM_dwarf_UBVIJHK_colors_Teff.txt} by Eric Mamajek. Version 2022.04.16.}, and applied that fit to our $G_{BP}-G_{RP}$ colors to compute \teff. Errors were propagated from the color calculation and RMS scatter of the polynomial fit, which yielded a median 2.93\% \teff\ uncertainty.

For all Kepler and K2 targets, we computed colors ($G_{BP}-G_{RP}$, $G-G_{RP}$, $G_{BP}-G$, $G_{RP}-J$, $J-H$, $H-K$) and absolute magnitudes ($M_J$, $M_H$, $M_{K_S}$), accounting for interstellar extinction using the \texttt{dustmaps} code \citep{Green2018} and the 3D Bayestar19 dust map \citep{Green2019}. We identified 63,009 Kepler and 73,038 unique K2 targets with a LAMOST DR9 v2.0 spectrum \citep{Cui2012} and [Fe/H] measurements. We calculated photometric [Fe/H] for all stars using random forest regression \citep[within \texttt{scikit-learn,}][]{Pedregosa2011}. We trained our random forest regressor on a random subset of 75\% of these spectroscopic targets and their associated colors and absolute magnitudes. The remaining 25\% of targets were used to check the reliability of the machine learning algorithm. The $R^2$ score of the trained random forest regressor was 0.55, indicating a moderately reasonable ability of the algorithm to predict metallicity. The resultant RMS scatter in [Fe/H] predicted versus measured values compared to a 1:1 line is 0.22 dex, which we adopt as our uncertainty in the predicted measurements of these values for our targets.

We computed bolometric luminosity from the bolometric magnitude, applying a bolometric correction calculated using \texttt{isoclassify} \citep{Huber2017} and the MIST stellar grid \citep{Choi2016}. We computed $R_{\star}$ using the Stefan-Boltzmann law and our bolometric luminosity and \teff\ values. For targets within the range $4.5 < M_{K_S} < 10$, we computed $R_{\star}$ using the empirical radius--luminosity--metallicity relationship for M dwarfs from \citet{Mann2015}. Target $M_{\star}$ was computed using a mass-luminosity relationship from \citet{Torres2010}, except for stars within the range $4.5 < M_{K_S} < 10$, for which we used the M dwarf empirical mass--luminosity--metallicity relationship from \citet{Mann2019}. Finally, we computed $\log g$ using Newton's law of universal gravitation and our $M_{\star}$ and $R_{\star}$ values. The median $R_{\star}$, $M_{\star}$, and $\log g$ uncertainties for our sample are 7.42\%, 8.15\%, and 1.81\%, respectively.

\begin{figure*}[ht!]
\centering
\includegraphics[width=0.325\linewidth,trim=2 2 2 2, clip]{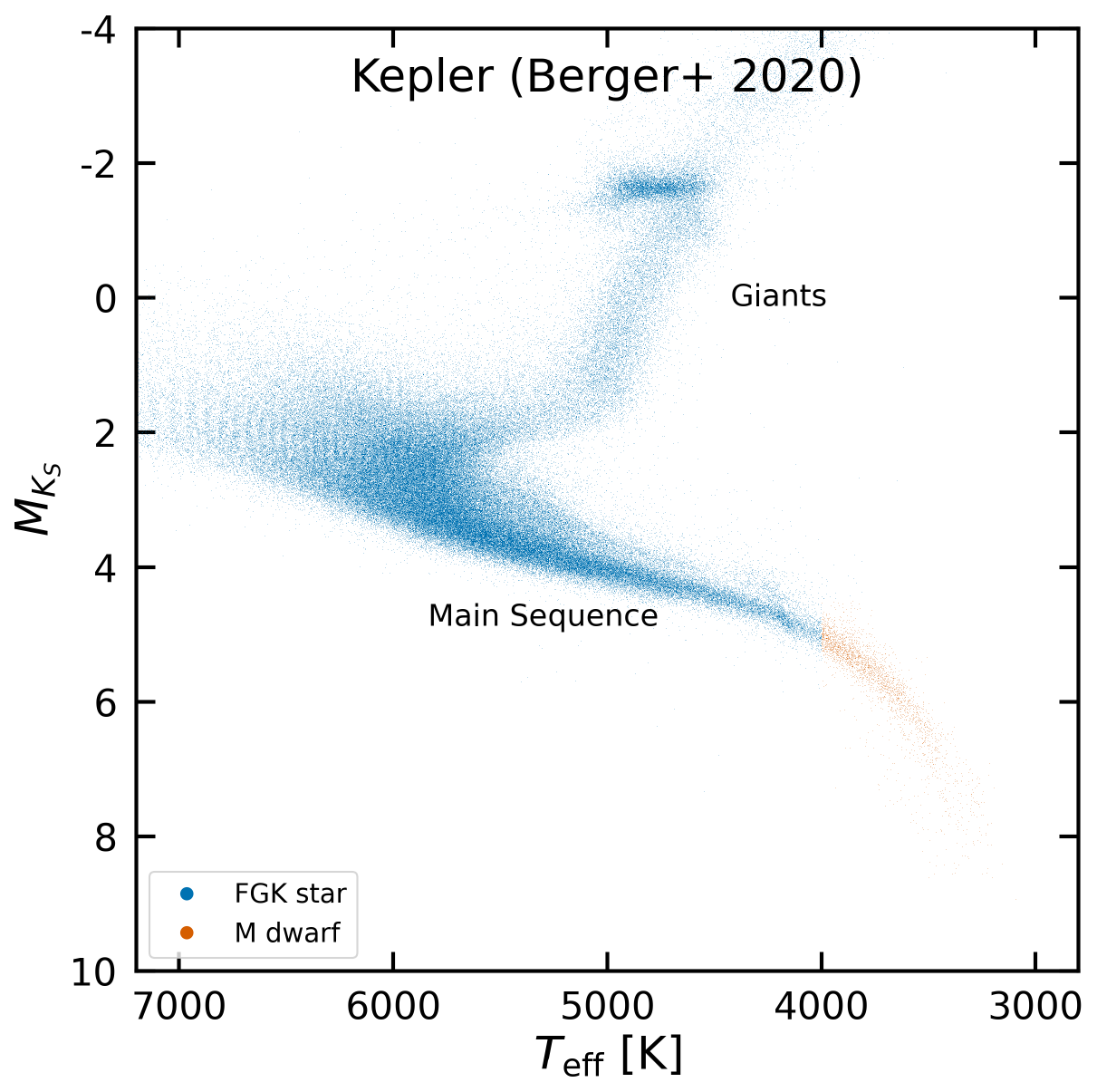}
\includegraphics[width=0.325\linewidth,trim=2 2 2 2, clip]{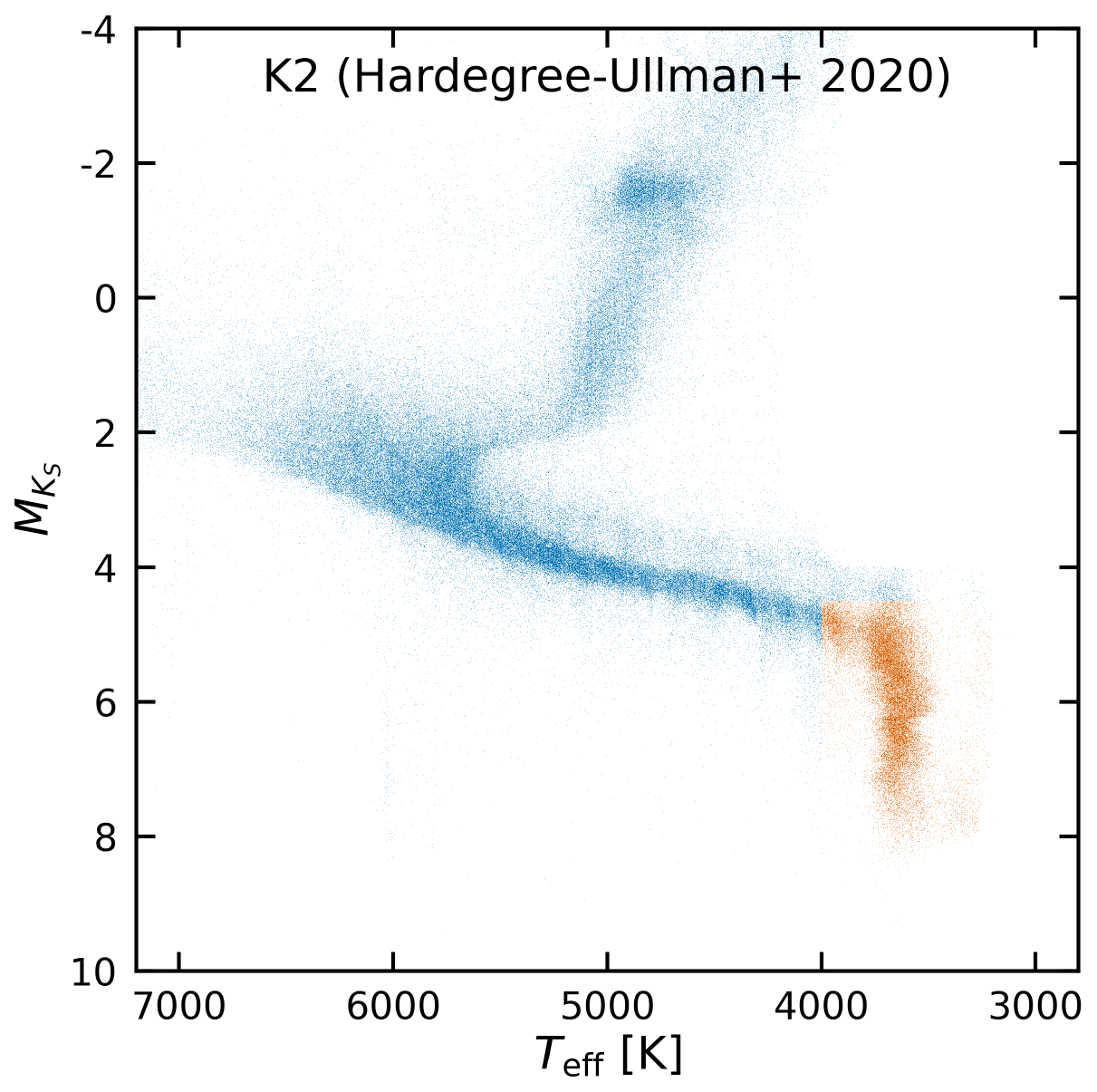}
\includegraphics[width=0.325\linewidth,trim=2 2 2 2, clip]{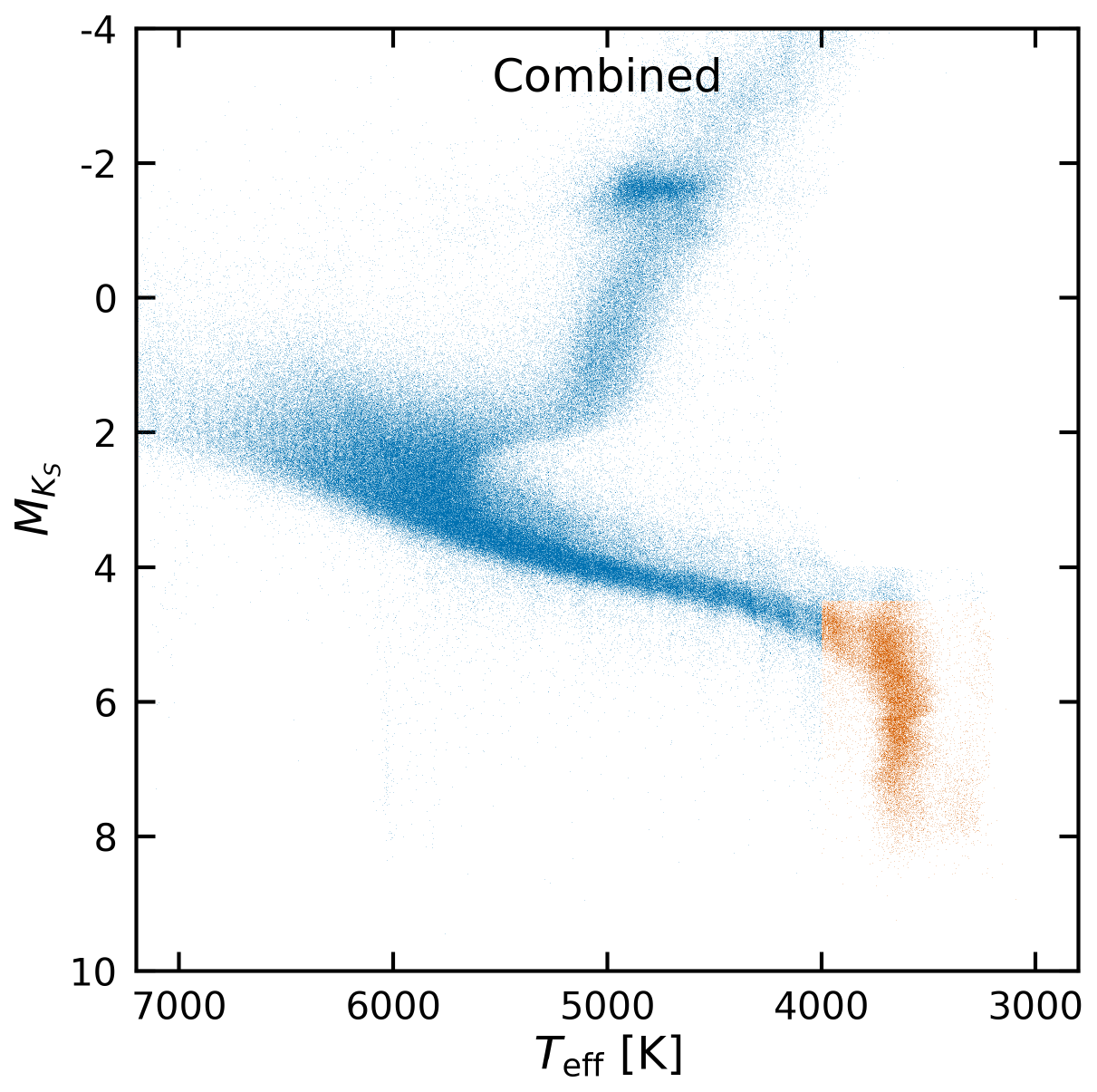} \\
\includegraphics[width=0.325\linewidth,trim=2 2 2 2, clip]{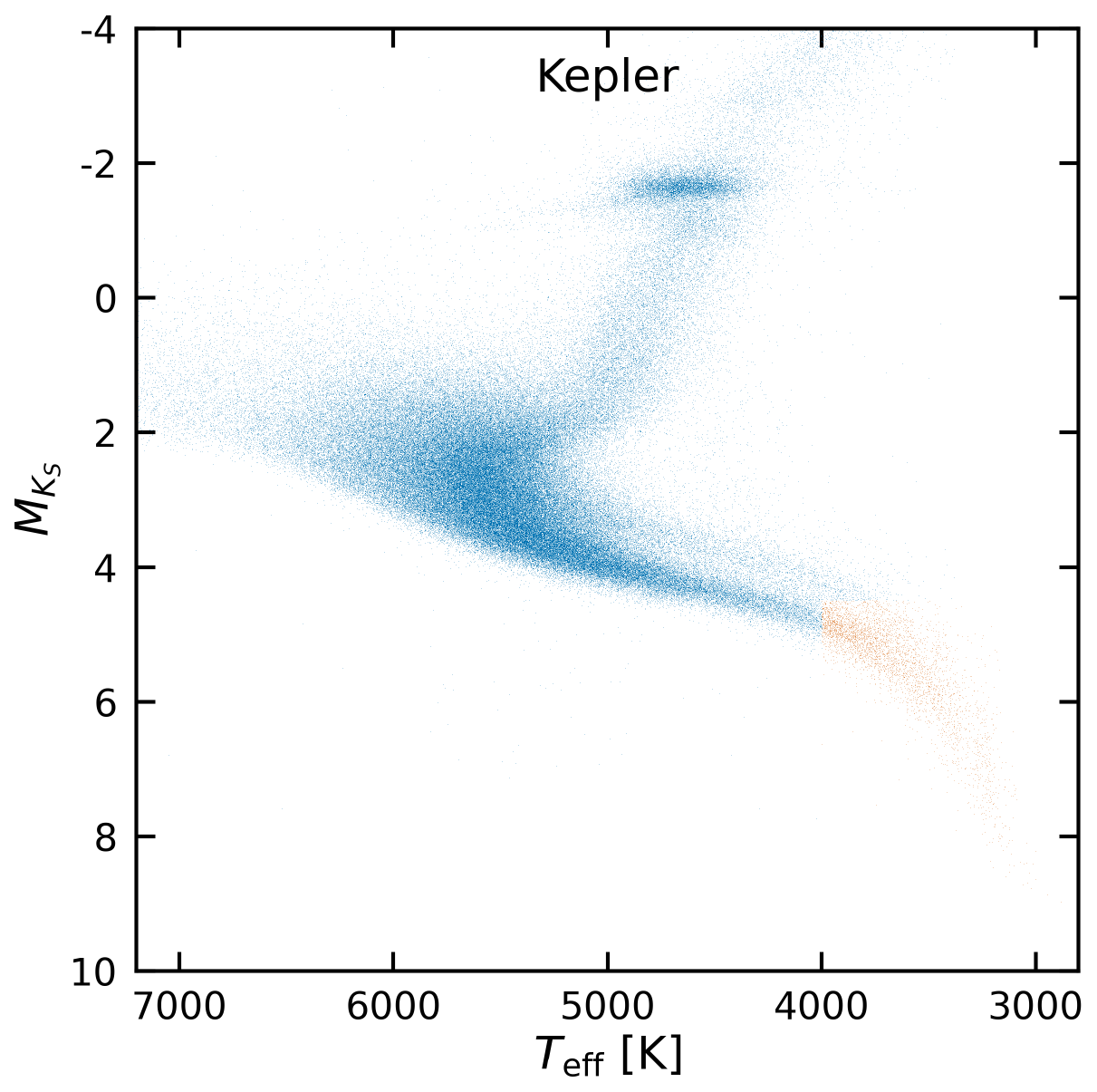}
\includegraphics[width=0.325\linewidth,trim=2 2 2 2, clip]{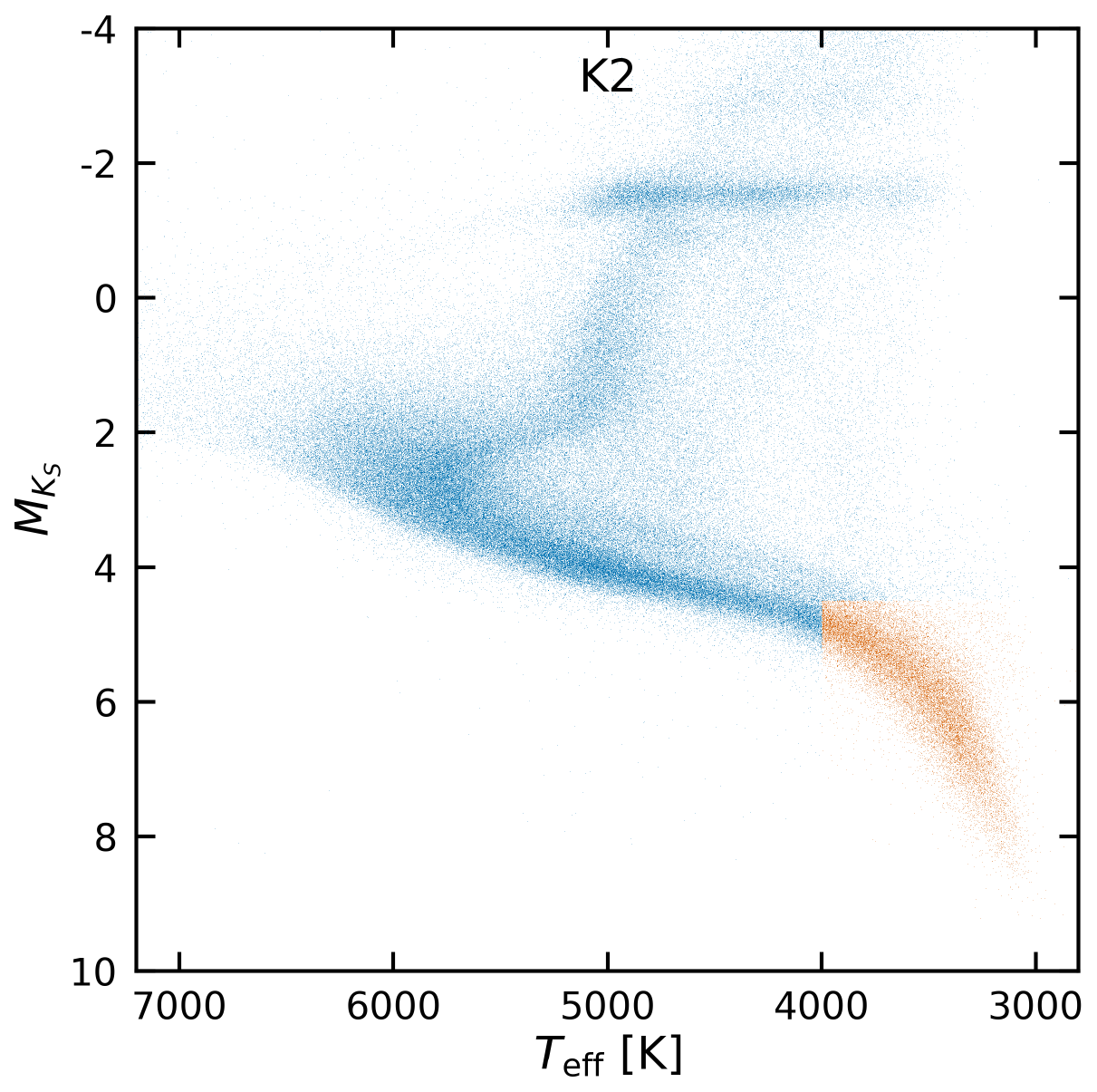}
\includegraphics[width=0.325\linewidth,trim=2 2 2 2, clip]{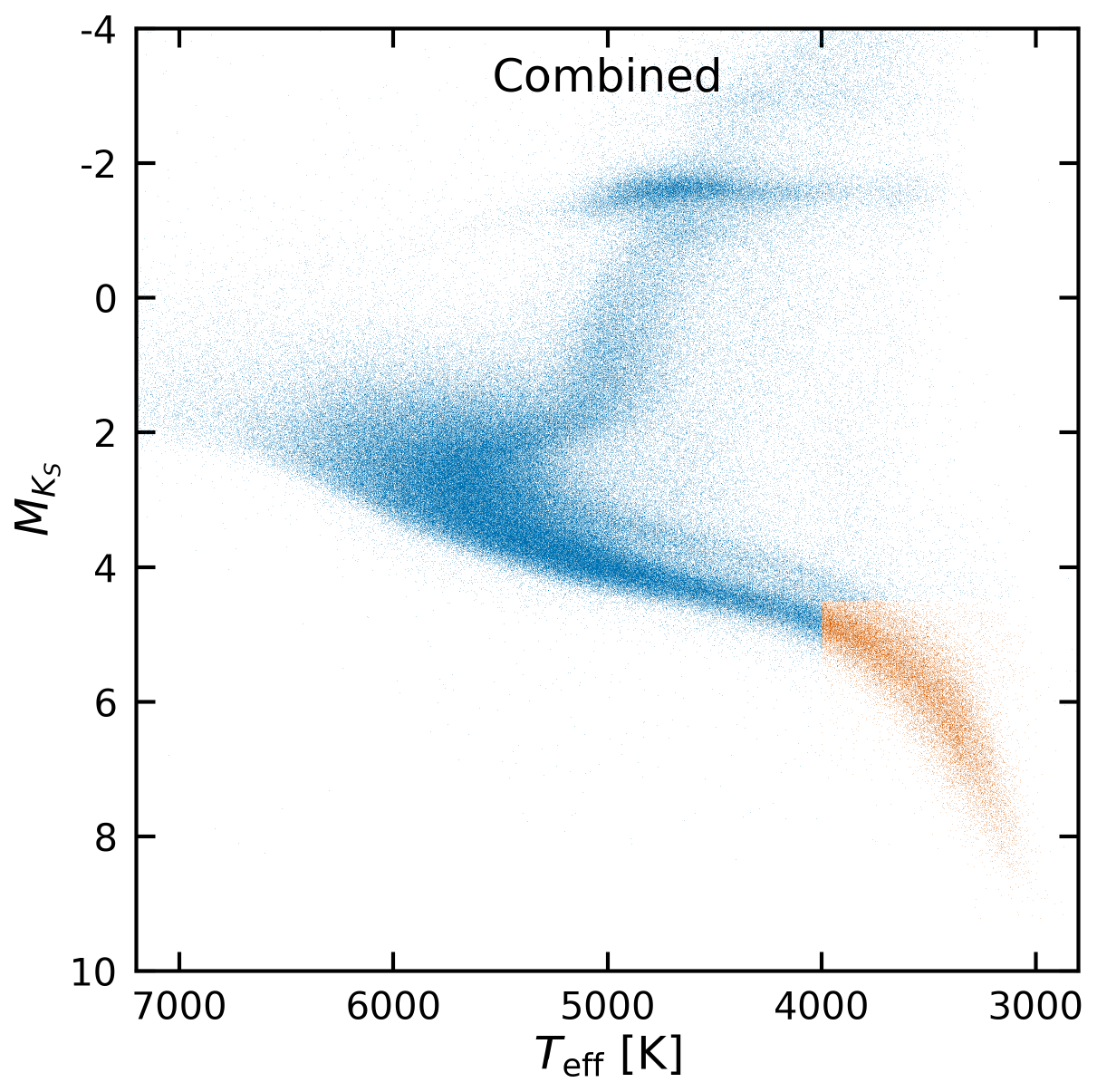}
\caption{Hertzsprung-Russell diagrams of the Kepler (left), K2 (middle), and combined Kelper and K2 stellar populations (right) from \citet{Berger2020b} and \citet{Hardegree-Ullman2020} (top) and our new uniform stellar classification (bottom). Stars in orange are M dwarfs based on temperature and absolute magnitude. The top panels highlight the original stellar parameter differences due to different methodologies and target populations. Most of the original K2 M dwarfs from \citet{Hardegree-Ullman2020} have temperatures near 3700~K due to the input training set limitations, which is mitigated with the new temperature derivation. It is clear that the Kepler mission was focused on Sun-like main sequence stars, but the guest-driven K2 mission more broadly sampled a wider population of F, G, K, and M stars. K2 notably observed nine times more M dwarfs than Kepler.} \label{fig:hr}
\end{figure*}

Hertzsprung-Russell diagrams of the Kepler and K2 targets are plotted in Figure~\ref{fig:hr}, showing the differences between the new uniform sample and previous catalogs from \citet{Berger2020b} and \citet{Hardegree-Ullman2020}. It is clear from these H-R diagrams that K2 observed significantly more M dwarfs. Using a simple $T_{\mathrm{eff}}<4000$\,K, $R_{\star}<0.6 R_{\star}$, and $M_{\star}<0.6 M_{\star}$ cut, K2 observed 9.34 times more M dwarfs than Kepler.

To further refine our stellar sample for comparisons to demographic studies of FGKM dwarfs, we only included stars with $T_\mathrm{eff}$ in the range of $2300-7300$~K and $\log g$ in the range of non-evolved stars based on the dwarf criterion from \citet{Huber2016} (their Equation 9), leaving 108,019 Kepler stars and 133,848 unique K2 stars. We also excluded probable binary stars with Gaia \texttt{RUWE} values greater than 1.4 and Gaia \texttt{non\_single\_star} not equal to $0$, leaving 95,092 Kepler stars and 113,402 unique K2 stars. Targets with noisy photometry provide little constraining power and merely slow down the calculations, therefore, we removed targets with $\textrm{CDPP}_\textrm{8hr}>1200$~ppm for K2 stars, as was done in previous Scaling K2 papers \citep{Zink2020a,Zink2020b,Zink2023,Christiansen2023}, and $\textrm{CDPP}_\textrm{7.5hr}>1000$~ppm for Kepler stars. These cuts left 94,422 Kepler and 90,344 unique K2 FGKM main sequence stars. Properties for the entire Kepler and K2 stellar samples are given in Appendix~\ref{app:A} in Tables~\ref{tab:kep} and \ref{tab:k2}. 

\section{Planet Sample} \label{sec:planets}

The planet sample is a combination of both Kepler and K2 planet candidates, which represent a homogeneously derived sample of transiting planets in the local galactic neighborhood. K2 planets were selected from the full Scaling K2 planet catalog in \citet{Zink2021} and Kepler planets were drawn from \citet{Berger2020a} based on Kepler DR25 \citep{Thompson2018}. Since the K2 stellar radii, masses, and temperatures were updated from the \citet{Hardegree-Ullman2020} sample (as discussed in Section \ref{sec:sample}), the planet radii and stellar incident flux were updated accordingly. This ensured consistency between the stellar and planet samples, culminating in a median planet radius and incident stellar flux change of 2.2\% and 11.7\%, respectively. The same updates were applied to the Kepler stellar and planet samples from \citet{Berger2020a} and \citet{Berger2020b}, leading to average planet radius and incident stellar flux changes of 2.3\% and 9.3\%, respectively. The refined planet parameters are given in Appendix~\ref{app:A} in Tables~\ref{tab:kepplanet} and \ref{tab:k2planet}. 

\begin{figure}[ht!]
\centering
\includegraphics[width=\linewidth]{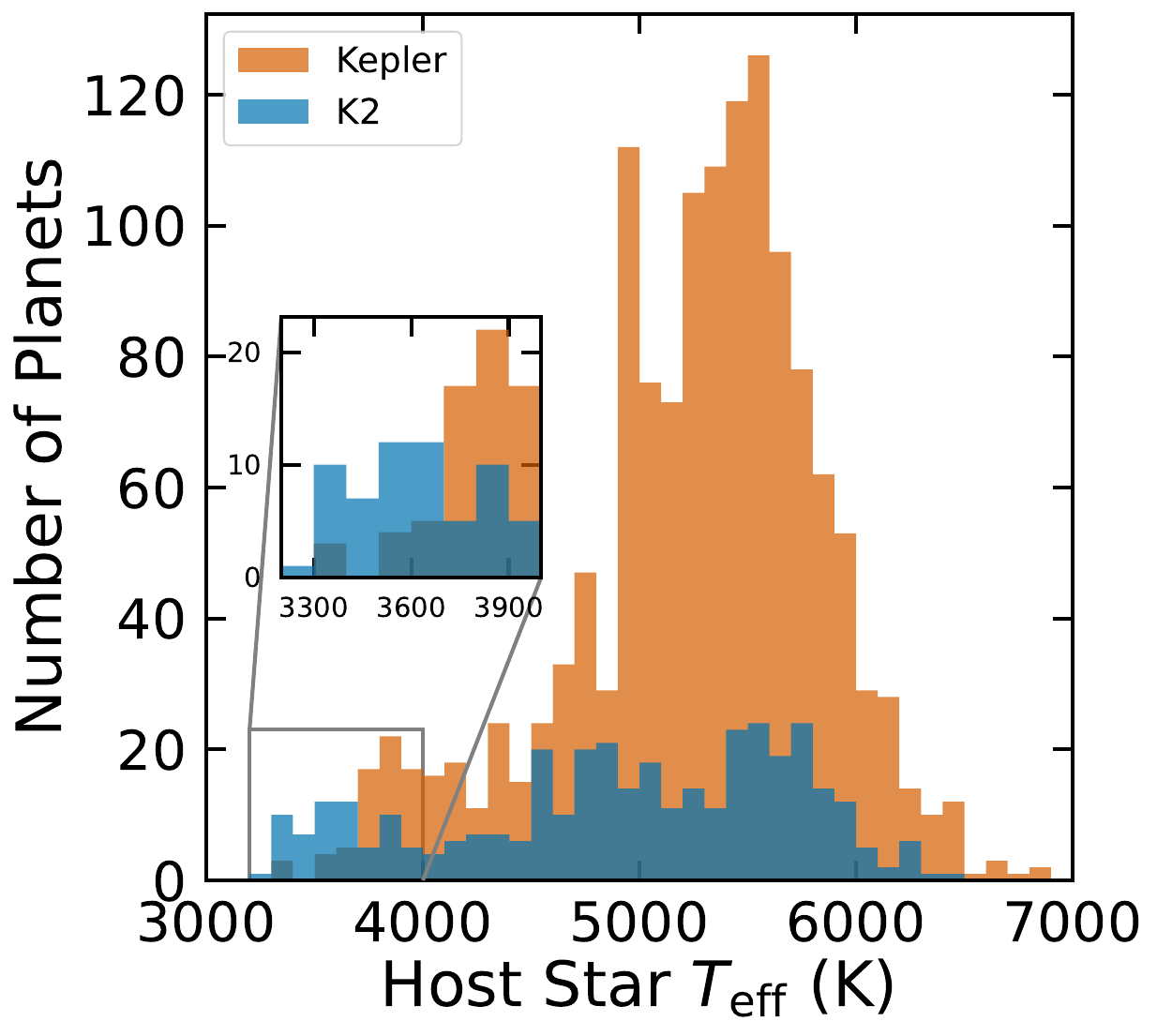}
\caption{Histograms of Kepler (orange) and K2 (blue) super-Earth and sub-Neptune host star effective temperatures used in this study. While comparable total numbers of super-Earth and sub-Neptune candidates were found orbiting M dwarfs ($T_{\mathrm{eff}}\lesssim4000$\,K) in both Kepler and K2, it is worth noting that for $T_{\mathrm{eff}}<3700$\,K, there are 3.5 times more K2 candidates than Kepler candidates (see inset plot).} \label{fig:teff-pl}
\end{figure}

The focus of this study is on super-Earths and sub-Neptunes, thus for the following analysis we only include planets with radii between 1 and 4 $R_\Earth$. Furthermore, we limit our analysis to planets with orbital period between 1 and 40 days. Beyond this period range, the K2 observing baseline does not permit detection of three transits. Additionally, we required transiting impact parameters $b\leq0.9$, minimizing the radius inaccuracies identified for near-grazing transits \citep{Gilbert2022}. These cuts left us with 1385 Kepler planets and 328 unique K2 planets with host stars between 3200~K and 6900~K. Figure~\ref{fig:teff-pl} shows a histogram of the number of planets for different host stellar temperatures. There are 68 Kepler and 62 K2 super-Earths and sub-Neptunes orbiting stars with $T_{\mathrm{eff}}<4000$. Most of the Kepler planets, however, are around earlier-type M dwarfs. For stars with $T_{\mathrm{eff}}<3700$, there are 12 Kepler and 42 K2 super-Earths and sub-Neptunes. This highlights the importance of K2 in constraining super-Earth and sub-Neptune planet occurrence rates for mid-type M dwarfs.

\begin{figure*}[htb!]
\centering
\includegraphics[height=9cm]{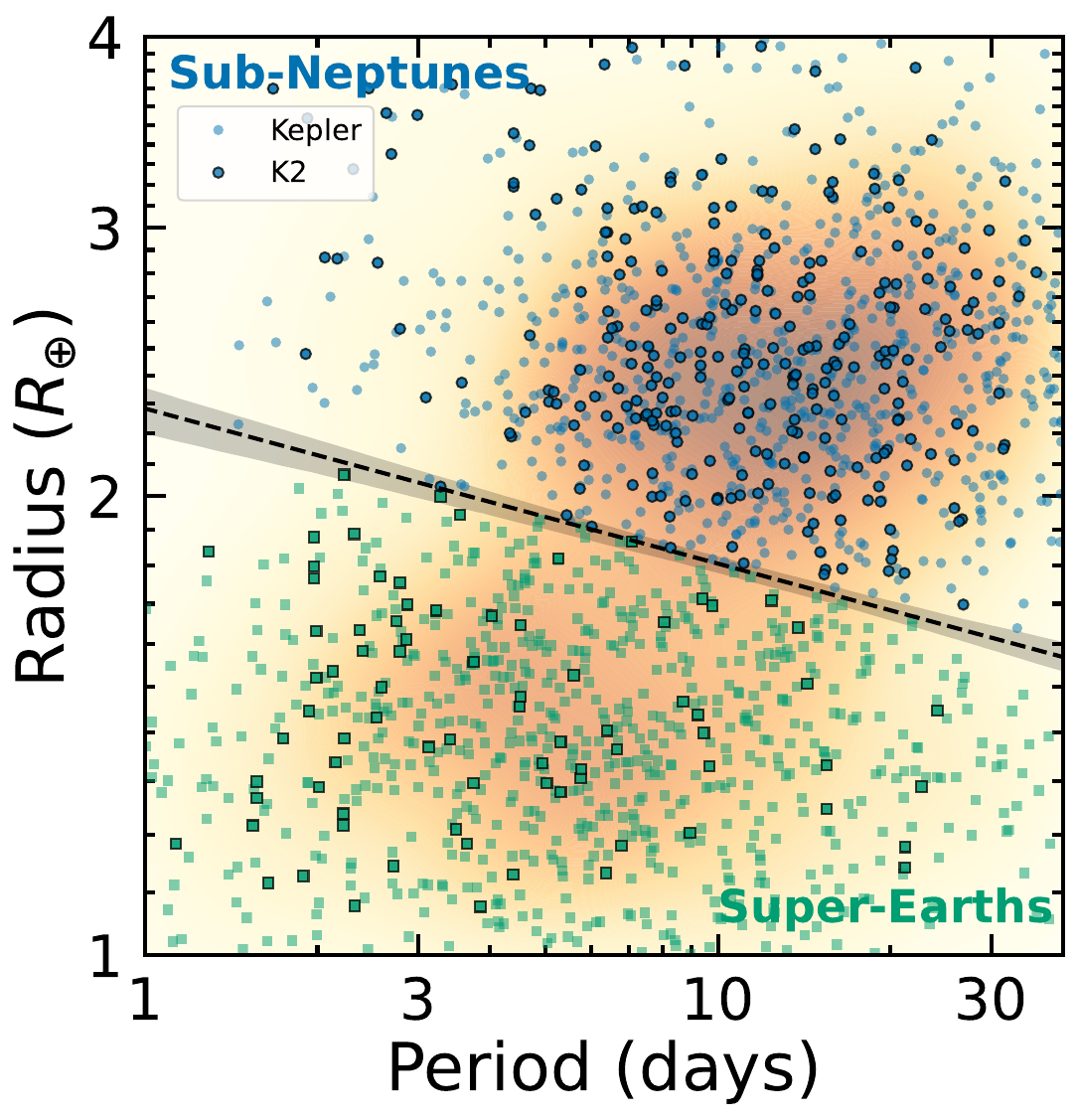}
\includegraphics[height=9cm]{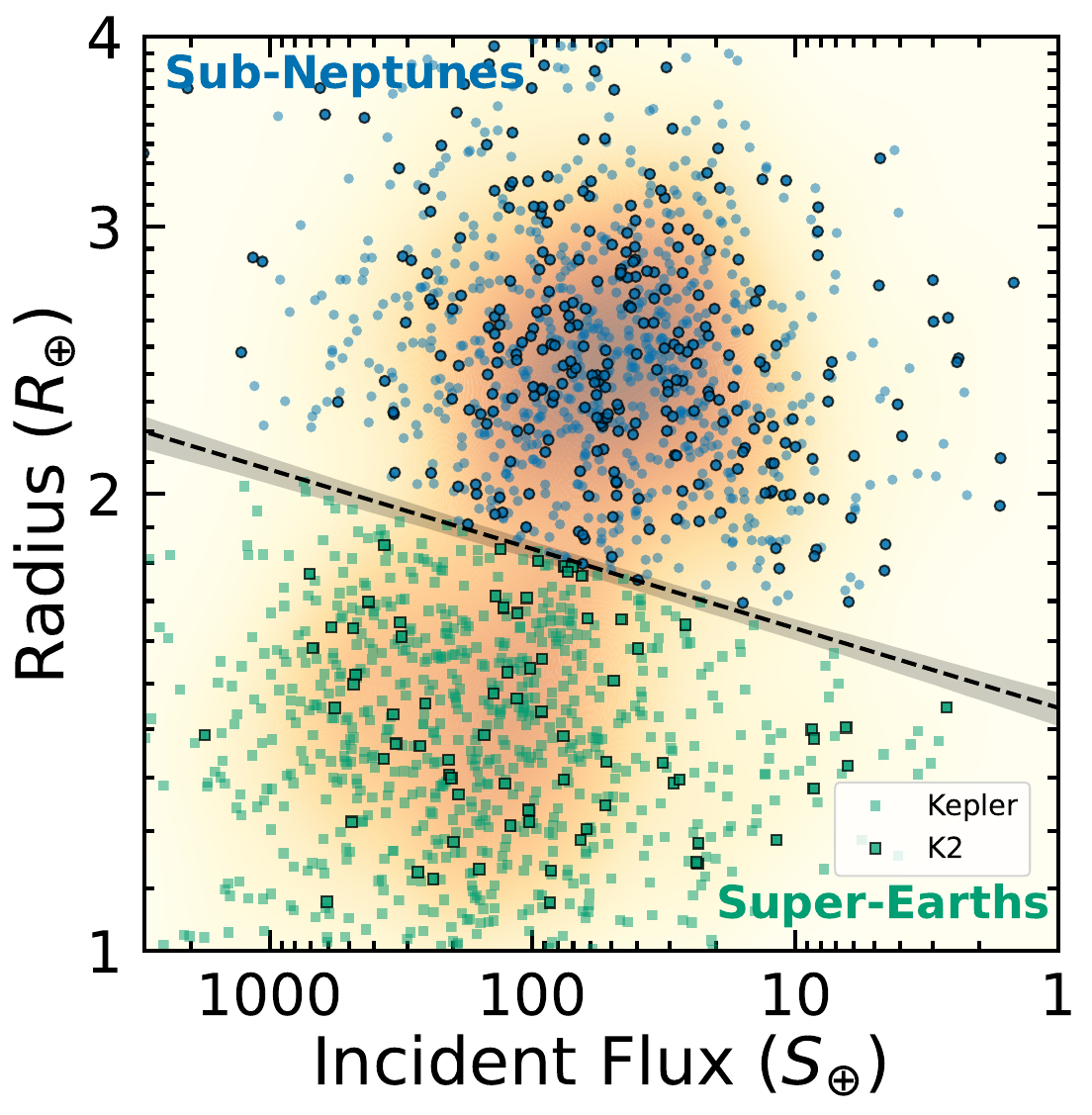}
\caption{(Left) Period and radius diagram and (right) incident flux and radius diagram of the combined Kepler (no outline) and K2 planet candidates (black outlined points) orbiting FGKM stars used for our analysis. The dotted lines are the fit to the partition between super-Earths and sub-Neptunes from a KDE fit to the radius valley (Equation~\ref{eq:radper}), with the shaded region highlighting the uncertainties from a 1000 bootstrap resampling of the planet population. Super-Earths are shown as green squares below the lines, and sub-Neptunes are shown as blue circles above the lines. The dark shaded regions behind the points highlight the density of exoplanets in the respective phase space. We note that we have not corrected for detection completeness in these plots. The corresponding parameters of the planet sample are provided in Tables~\ref{tab:kepplanet} and \ref{tab:k2planet}.} \label{fig:planet}
\end{figure*}

The exoplanet radius valley \citep{Fulton2017,VanEylen2018,Hardegree-Ullman2020} provides a natural separation between thin enveloped super-Earths and thick H/He enveloped sub-Neptunes. The exoplanet radius valley appears to have a slope in orbital period or incident-flux space \citep[e.g.,][]{VanEylen2018, Petigura2022, Ho2023}, and is likely the outcome of some stellar-driven mass loss mechanism (photoevaporation; \citealt{Owen2017}, or core-powered mass-loss; \citealt{Ginzburg2018}). Therefore, we used Gaussian kernel density estimation (KDE) to fit the exoplanet radius valley for FGKM stars in the form:
\begin{equation}\label{eq:radper}
    \log_{10} \frac{R_p}{R_{\oplus}} = m \log_{10}\frac{P}{\mathrm{days}} + y_0,
\end{equation}
where $m=-0.102^{+0.014}_{-0.011}$ and $y_0=0.359^{+0.012}_{-0.017}$. Modifying Equation~\ref{eq:radper} to exchange period for incident stellar flux ($S$), we fit the radius valley for FGKM stars with $m=0.160^{+0.010}_{-0.012}$ and $y_0=0.052\pm0.005$. The slope values are consistent with the fits by \citet{Petigura2022} to FGK and early-type M dwarf hosts down to 0.5~$M_\odot$ and \citet{Ho2023} to FGK stellar hosts. We plot the combined Kepler and K2 planet populations in radius--period and radius--incident flux space in Figure~\ref{fig:planet}, along with a fit to the radius valley. Using these fits to the radius valley, we can identify targets as either super-Earths if they fall below the line, and sub-Neptunes if they fall above the line. Only 2.3\% of targets have inconsistent super-Earth/sub-Neptune dispositions between period and incident flux space. These targets are all very near the radius valley fit in both scenarios. Better constraints on orbital period and incident flux could further minimize these inconsistencies.

\section{Planet Occurrence Rates} \label{sec:rates}

Given a planet and host star sample, meaningful occurrence trends can be extracted. This requires a clear and quantitative understanding of the inherent planet sample biases. In Kepler \citep{Christiansen2020} and K2 \citep{Zink2021}, injection/recovery tests were carried out against their respective automated search algorithms. Using the FGKM reliability maps from \citet{Thompson2018} (Kepler) and \citet{Zink2021} (K2), the \texttt{ExoMult} \citep{Zink2019} forward modeling software can simulate the planet population throughput. Within \texttt{ExoMult}, completeness is uniquely calculated for each bin of stellar effective temperature and separately for Kepler and K2, where we simulate detection efficiency for each star of relevant temperature, along with subsample-specific calculations of the geometric transit probability. This approach captures how the different completeness elements vary with stellar temperature, allowing us to properly account for changes in the completeness corrections (and thus occurrence rates) with temperature. An initial planet population model is passed through \texttt{ExoMult} and the simulated output population parameters are compared against the observed planet population parameter via a collection of Anderson-Darling tests and a modified Poisson likelihood. The initial planet population model is then modulated numerous times until the test statistic is appropriately optimized. For a more in-depth discussion of the optimization process and the underlying test statistic, see \citet{Zink2020b,Zink2023}.

The current study focuses on the effects of stellar $T_\text{eff}$ on planet occurrence rates, thus we follow the methodology of \citet{Zink2023} and use the following power law to model the underlying planet population:

\begin{equation} \label{eq:pl}
\frac{dn}{d T_{\text{eff}}} = \eta \cdot \left(\frac{T_\text{eff}}{5000\,\mathrm{K}}\right)^\gamma
\end{equation}
where n is the number of planets (super-Earths, sub-Neptunes, or both), $\eta$ represents the normalization of planets and $\gamma$ is a tunable parameter.

In addition, planet radius and period distributions can significantly impact the inferred model parameters. To best represent the underlying planet period and radius distributions, we simulated planets drawn from the best-fit combined Kepler and K2 model from \citet{Zink2023}. Briefly, sub-Neptunes and super-Earths were separately drawn from a power law in radius space and a broken power law in orbital period space, which has been used in numerous exoplanet occurrence rate studies \citep{Youdin2011,Petigura2013,Mulders2018,Zink2019,Bergsten2022,Zink2023}.  The occurrence rate model in period and radius space is then:

\begin{equation}
    \frac{d^2n}{d\log P_{\mathrm{pl}}\ d\log R_{\mathrm{pl}}} = f\ g(P_{\mathrm{pl}})\ q(R_{\mathrm{pl}}),
\end{equation}
where $n$ is the number of planets, $P_{\mathrm{pl}}$ is orbital period, $R_{\mathrm{pl}}$ is planet radius, $f$ is a normalization factor, and $g(P_{\mathrm{pl}})$ and $q(R_{\mathrm{pl}})$ are the planet period and radius power laws defined as:
\begin{equation}
    \begin{split}
        g(P_{\mathrm{pl}}) & \propto
        \begin{cases}  
            P_{\mathrm{pl}}^{\beta_1}\ \ \mathrm{if}\ P_\mathrm{pl}< P_\mathrm{br}\\
            P_{\mathrm{pl}}^{\beta_2}\ \ \mathrm{if}\ P_\mathrm{pl} \geq P_\mathrm{br}\\
        \end{cases}\\
        q(R_\mathrm{pl}) & \propto R_\mathrm{pl}^\alpha,
    \end{split}
\end{equation}
where $\beta_1$, $\beta_2$, and $\alpha$ are the scaling model parameters, and $P_\mathrm{br}$ is the break in the period power law. The optimized model parameters we use from \citet{Zink2023} are listed in Table~\ref{tab:optmod}.

\begin{deluxetable*}{cccccc}
\tablecaption{Optimized Combined FGKM Kepler and K2 Exoplanet Occurrence Rate Model for Orbital Period and Radius Space \label{tab:optmod}}
\tablehead{\colhead{Class} & \colhead{$f$} & \colhead{$\beta_1$} & \colhead{$\beta_2$} & \colhead{$P_\mathrm{br}$} & \colhead{$\alpha$}} 

\startdata
Super-Earth & 0.31$\pm$0.02 & 2.0$\pm$0.2 & 0.2$\pm$0.2 & 5.6$\pm$1.1 & -1.1$\pm$0.2 \\
Sub-Neptune & 0.30$\pm$0.01 & 2.5$\pm$0.4 & 0.5$\pm$0.2 & 9.5$\pm$2.0 & -1.7$\pm$0.2 \\
\enddata

\end{deluxetable*}

After visual inspection, we found that a single power law in stellar effective temperature, as given in Equation~\ref{eq:pl}, was not sufficient to model the underlying population of sub-Neptunes. Rather a double broken power law of the following form was used:

\begin{equation} \label{eq:dbpl}
\begin{split}   
\frac{dn}{d T_{\text{eff}}} =
\begin{cases}  
\eta \cdot \left(\frac{T_\text{eff}}{T_\text{Br,1}}\right)^\xi & \text{if $T_\text{eff} < T_\text{Br,1}$}\\
\eta \cdot \left(\frac{T_\text{eff}}{T_\text{Br,1}}\right)^\gamma & \text{if $T_\text{Br,1} < T_\text{eff} < T_\text{Br,2}$}\\
\eta \cdot \left(\frac{T_\text{Br,2}}{T_\text{Br,1}} \right)^\gamma \left(\frac{T_{\text{eff}}}{T_\text{Br,2}} \right)^\zeta & \text{if $T_{\text{eff}} > T_\text{Br,2}$},\\
\end{cases}
\end{split}
\end{equation}
where $T_\text{Br,1}$ and $T_\text{Br,2}$ are the effective temperatures of the power law breaks, and $\gamma$ and $\zeta$ are additional tuning parameters. We also attempted to fit the super-Earth population with a broken power law; however, the break parameter was very close to the upper boundary of our temperature range and thus was not well constrained. Hence, we stuck with a single power law for the super-Earth population.
 
\begin{figure*}[ht!]
\centering
\includegraphics[width=\linewidth]{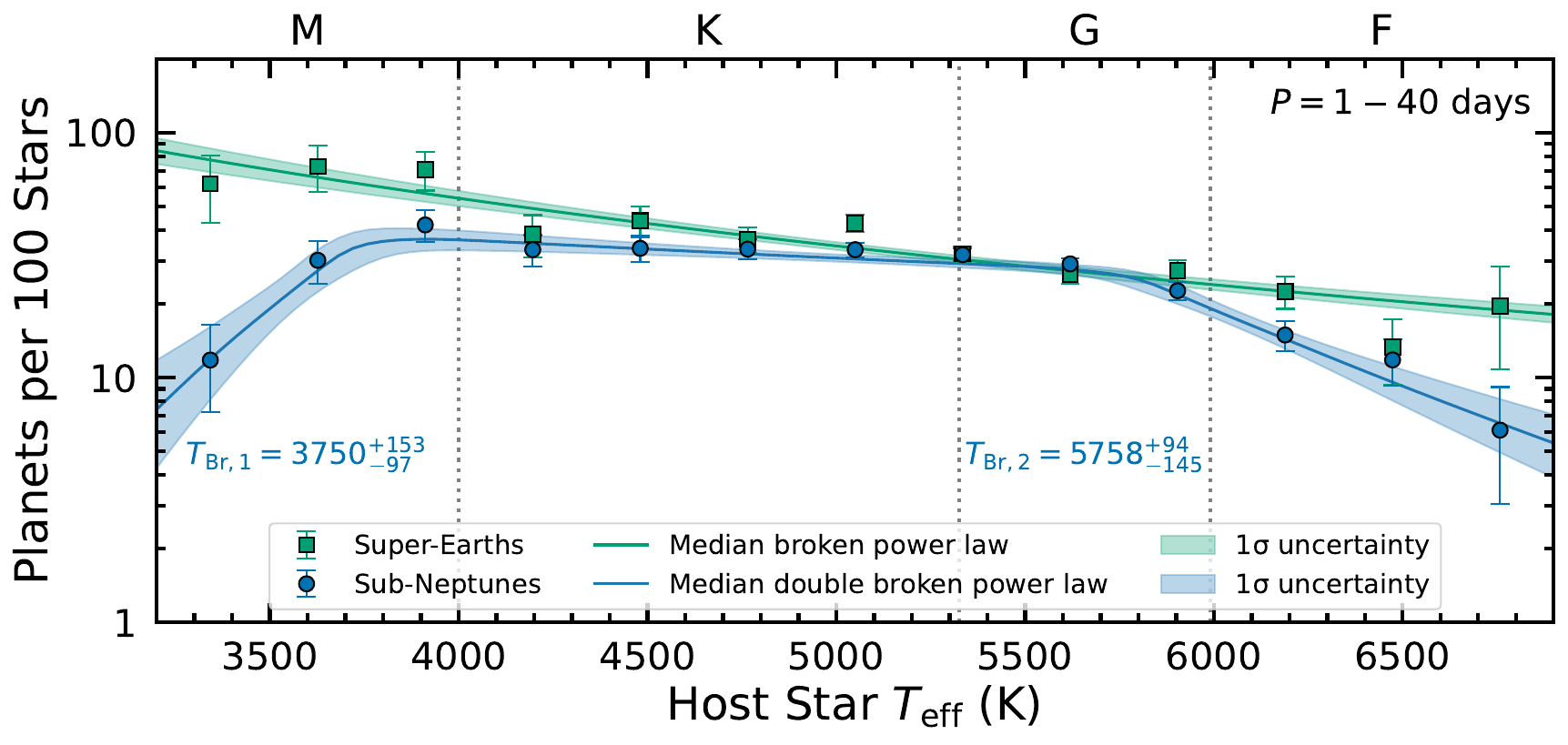}
\caption{Planet occurrence rates as a function of stellar temperature for short-period ($P=1-40$~days) super-Earths (green squares) and sub-Neptunes (blue circles). A power law was fit to the super-Earths, and a double broken power law was fit to the sub-Neptunes, showing two breaks: $T_{\mathrm{Br,1}}=3750^{+153}_{-97}$~K and $T_{\mathrm{Br,2}}=5758^{+94}_{-145}$~K. The short-period sub-Neptunes peak around a host type of M0.5~V, then decrease in planet occurrence toward cooler M dwarfs. G and K-type stars have comparable rates of short-period super-Earths and sub-Neptunes. Another noticeable decrease in sub-Neptunes happens for host stars hotter than the Sun. Short-period super-Earths increase in occurrence toward cooler stars, with no discernible peak in the host star temperatures for which we have data from Kepler and K2.} \label{fig:occ}
\end{figure*}

The respective population parameters ($\eta$, $\alpha$, $\gamma$, and $\zeta$) were varied over the course of 100,000 simulations, using uniform priors. The resulting parameter posteriors are displayed in Table \ref{tab:dbpl}, and the planet occurrence rates are shown in Figure~\ref{fig:occ}.  It is worth noting that for the sub-Neptunes with orbital periods shorter than 40 days, the two power law breaks were identified at $3750^{+153}_{-97}$~K and $5758^{+94}_{-145}$~K. We discuss the implications of these results in the following section.

\begin{deluxetable*}{ccccccc}
    \tablecaption{Occurrence Rate Power Law Parameters\label{tab:dbpl}}
    \tablehead{\colhead{Planet Type} & \colhead{$\eta$} & \colhead{$T_\text{Br,1}$} & \colhead{$T_\text{Br,2}$} & \colhead{$\xi$} & \colhead{$\gamma$} & \colhead{$\zeta$} }
    
    \startdata
    Sub-Neptune & $0.38 \pm 0.04$ & $3750^{+153}_{-97}$~K & $5758^{+94}_{-145}$~K & $10.66^{+5.32}_{-4.33}$ & $-0.81^{+0.38}_{-0.36}$ & $-8.95^{+2.02}_{-2.53}$\\
    Super-Earth & $0.35\pm0.01$ & \nodata & \nodata & \nodata & $-2.01\pm0.24$ & \nodata \\
    \enddata

\end{deluxetable*}

\section{Discussion \& Future} \label{sec:discussion}

With this new homogeneous analysis combining Kepler and K2 data we were able to measure detailed occurrence rates of short-period super-Earths and sub-Neptunes orbiting M dwarfs with \teff\ between 3200~K and 4000~K. We find that short-period sub-Neptune occurrence rates peak around stars near host stars of spectral type M0~V and decrease at cooler temperatures (Figure~\ref{fig:occ}), which we discuss in the following subsection. Between M0~V and solar-temperature stars, there is a shallower decrease in short-period sub-Neptunes, followed by another steep decrease for stars hotter than the Sun. We leave the analysis and discussion of the hot star sub-Neptune drop to a future paper (Hardegree-Ullman et al. in prep). Additionally, there is no clear peak for super-Earths (Figure~\ref{fig:occ}, left panel), but there is a $>1\sigma$ ``jump'' at the M dwarf transition, followed by a potential plateau in occurrence rates.

\subsection{The M Dwarf Sub-Neptune Peak: Comparison to Planet Formation Models}\label{sec:models}

Our observational result of a peak in short-period sub-Neptunes loosely corresponds to recent theoretical planet formation models from \citet{Mulders2021}, \citet{Burn2021}, \citet{Chachan2023}, and \citet{Pan2025}. \citet{Mulders2021} simulated the formation of two-planet systems orbiting stars from 0.1--2.0~$M_{\odot}$ using a pebble accretion evolutionary disk model from \citet{Drazkowska2021}. They found that for the lowest-mass stars (0.1--0.2~$M_{\odot}$), no planet cores become massive enough to become giant planets, and only the most massive disks can form planets between 1 and 4~$R_{\oplus}$. For 1--2~$M_{\odot}$ stars, an outer giant planet suppresses pebble flux to the inner disk, inhibiting the formation of inner small planets. For 0.5~$M_{\odot}$ stars, a balanced pebble flux allows quick smaller planet formation and suppression of giant planet formation, causing a peak in 1--4~$R_{\oplus}$ planet occurrence. 

\citet{Burn2021} used a planet formation and evolution population synthesis model, accounting for planet migration, interactions, accretion, and atmospheric loss in their model. Their simulations showed that Earth-sized planets with orbits up to 500 days are most common around early-type M dwarfs, and less common for Sun-like stars and late-type M dwarfs. 

\citet{Chachan2023} combined pebble accretion models and measured protoplanetary disk masses in order to further study both why cooler stars have fewer giant and more small planets, and the correlation between inner super-Earths and outer giant planets. Their models suggest a drop in the fraction of disks that can support creation of planets up to $2 M_{\oplus}$ around stars $M_{\star}<0.3-0.5 M_{\odot}$. This leads to a drop in the fraction of sub-Neptunes, but not necessarily for rocky super-Earths toward later-type M dwarfs. \citet{Chachan2023} provide several different models of fraction of disks that can create inner planets using different initial conditions, but these models are not directly translatable to planet occurrence rates, so we chose not to include them in the comparison below.

\citet{Pan2025} modeled planet formation using a pebble accretion-based planet population synthesis model which included pre-main-sequence stellar luminosity evolution and accounted for the chemical compositions of pebbles at different regions within the protoplanetary disk. For warm (10--100 days) super-Earths (1--10~$M_{\oplus}$ or 1--2.8~$R_{\oplus}$), they found that planet occurrence rates peaked around a host star mass of 0.5~$M_{\odot}$.

\begin{figure*}[ht!]
\centering
\includegraphics[width=0.7\linewidth]{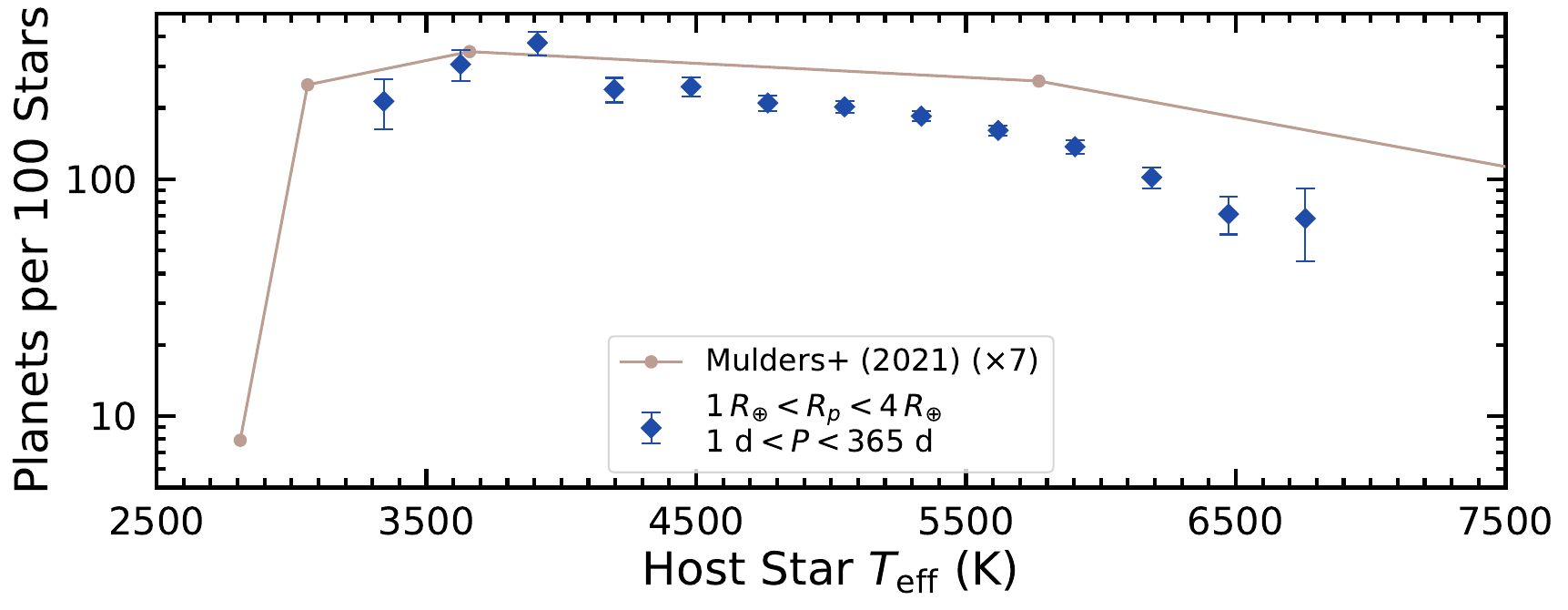}\\
\includegraphics[width=0.7\linewidth]{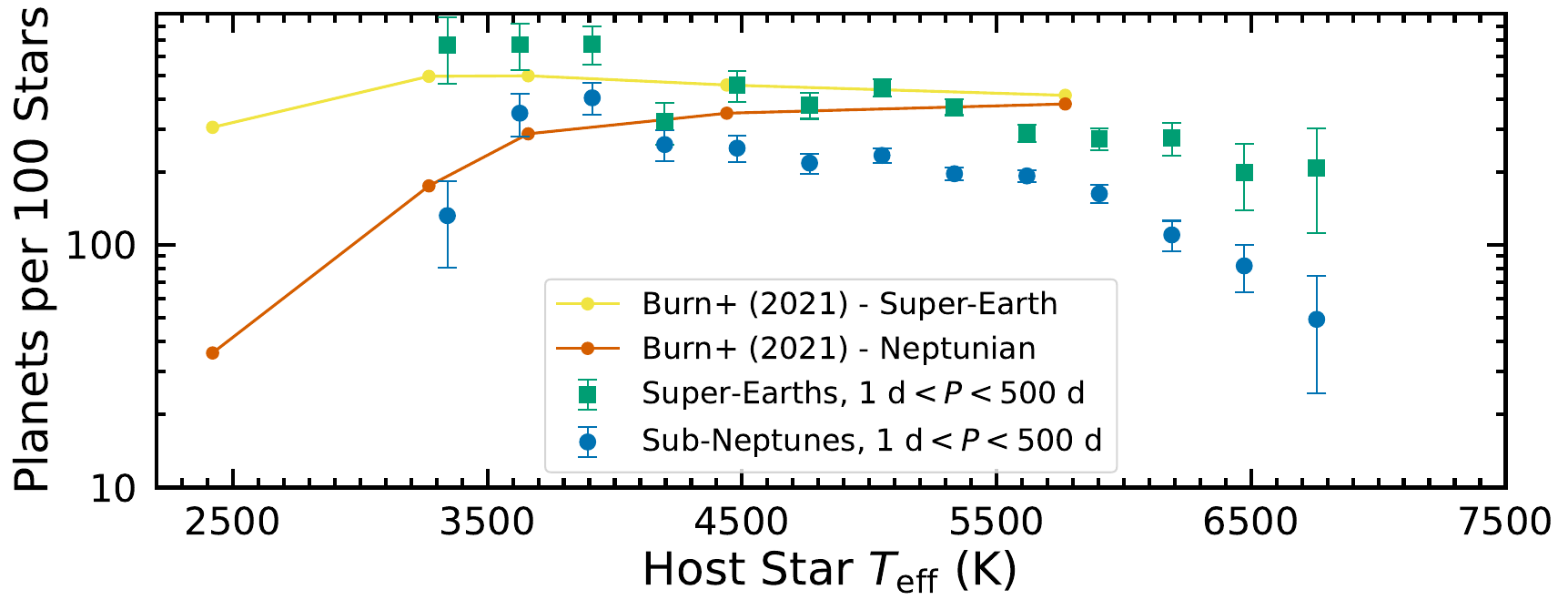}\\
\includegraphics[width=0.7\linewidth]{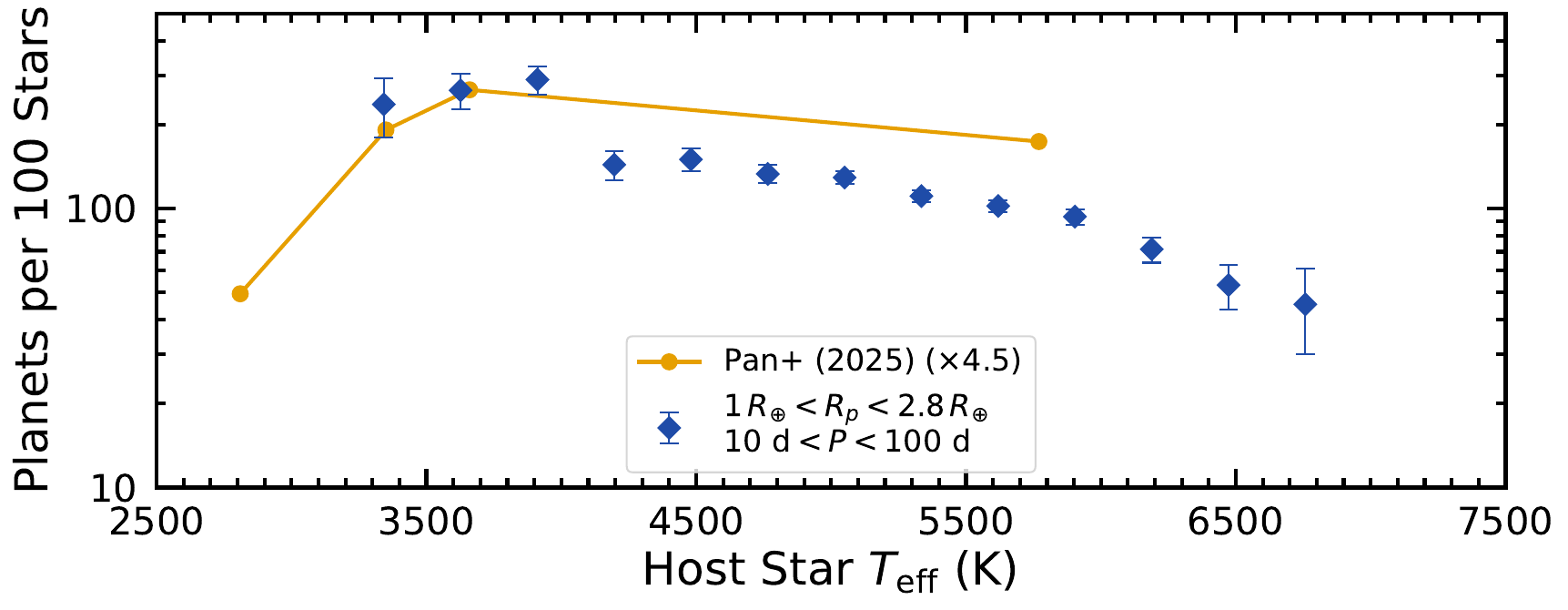}
\caption{Comparison of Kepler+K2 planet occurrence rates to formation model planet occurrence rates from (top panel) \citet{Mulders2021}, (middle panel) \citet{Burn2021}, and (lower panel) \citet{Pan2025}. Kepler+K2 occurrence rates were calculated at either the same, or as close as possible to the limits specified by each model. A scaling factor of 7 and 4.5 were applied to the models from \citet{Mulders2021} and \citet{Pan2025}, respectively, in order to roughly match the peak of the models to the peak in the K2+Kepler occurrence rates. The models were also originally in terms of host star mass and were converted to a host star effective temperature using an updated version of Table 5 from \citet{Pecaut2013}. Qualitatively, the models loosely follow the trends seen from observations. Our new planet occurrence rates can help guide further planet formation modeling efforts to more accurately reflect the observed population.} \label{fig:occ-models}
\end{figure*}

Figure~\ref{fig:occ-models} compares our planet occurrence rates to the planet formation-based occurrence rate models of \citet{Mulders2021}, \citet{Burn2021}, and \citet{Pan2025}. Since the models were given in terms of host star mass rather than temperature, we interpolated the values from updated Table~5 of \citet{Pecaut2013} to convert host mass to effective temperature. Comparing measured planet occurrence rates to these models, we note: (1) Unlike in Figure~\ref{fig:occ}, where planet sizes were limited to super-Earths and sub-Neptunes and an orbital period range between 1 and 40 days (the upper limit of orbital periods for K2 planets), each model has different boundaries as shown in Figure~\ref{fig:occ-models}. There are no K2 planets with orbital periods beyond 40 days, so the occurrence rates are reliant on the Kepler data there. Consequently, it is not a robust comparison for the M dwarfs, for which there are only 10 Kepler planets with orbital periods beyond 40 days, three of which have host star effective temperatures below 3750~K. (2) Each of the three models roughly match the location of the observational occurrence rate peak for small planets, but the behavior on either side of the peak is not fully consistent between models and observations. We have also applied a multiplicative offset in order to align the peaks in the \citet{Mulders2021} and \citet{Pan2025} models. (3) There is a distinct `jump' in the planet occurrence rates around 4000~K (roughly between K and M dwarfs). In all cases, the significance is $<3\sigma$.

Our new observational results provide important constraints for planet formation models. We encourage future modeling efforts to (1) distinguish between super-Earths and sub-Neptunes, since planet occurrence rates for these populations are distinct for M dwarfs, and (2) generate models on a finer stellar host grid to capture any other potential features, such as the sub-Neptune break near solar-type stars.

\subsection{Analysis of M Dwarf Planet Occurrence Rates}

While our sub-Neptune Kepler/K2 results are qualitatively consistent with recent planet formation models, it is unclear what happens to the super-Earth population toward stars cooler than $\sim$3400~K. Kepler studies identified that Earth- to Neptune-sized planets were more common toward smaller stars \citep{Mulders2015a,Mulders2015b,Dressing2015}. \citet{Hardegree-Ullman2019} also showed tentative evidence (with a sample of 13 planets) that this upward trend continued toward Kepler mid-type M dwarfs. However, the non-detection of planets orbiting late-type M dwarfs in planet occurrence rate studies casts observational doubt on a continued upward trend of small planets orbiting the smallest stars \citep{Sagear2020,Sestovic2020,Dietrich2023}. Further, \citet{Brady2022} and \citet{Ment2023} looked at the TESS mid and late-type M dwarf planet yields and found that planet occurrence rates likely do not increase toward the latest M dwarfs (though, with relatively small planet samples of 11 and 7, respectively). \citet{Brady2022} point out that TESS has detected less than half of the predicted planet yield for mid and late-type M dwarfs, but they note this could be attributed to an increase in photometric noise in the TESS bandpass for later M dwarfs. Ground-based searches for exoplanets orbiting nearby M dwarfs such as MEarth \citep{Irwin2009}, TRAPPIST \citep{Jehin2011}, SPECULOOS \citep{Delrez2018} and EDEN \citep{Gibbs2020} have so far yielded few planet discoveries for robust statistics. Notably, the discovery of the TRAPPIST-1 system \citep{Gillon2016} showed that high-multiplicity systems of rocky planets can form around the latest-type stars.

We ran additional calculations with \texttt{ExoMult} to compare planet occurrence rates for M dwarfs from the literature, including \citet{Dressing2015}, \citet{Mulders2015a}, \citet{Hardegree-Ullman2019}, and \citet{Ment2023}. The results from this comparison are given in Table~\ref{tab:occratecomp}. We generally attribute any significant differences to not using the exact same planet population (e.g., early vs early+mid vs mid+late M dwarf hosts), sample sizes, and differences in occurrence rate methodology (\texttt{ExoMult} forward modeling vs inverse detection efficiency method).

\begin{deluxetable*}{ccccc}
\tablecaption{Comparison to Literature M Dwarf Planet Occurrence Rates}\label{tab:occratecomp}

\tablehead{\colhead{Reference} & \colhead{Period} & \colhead{Radius} & \colhead{Planets per star} & \colhead{Planets per star}\vspace{-0.75em} \\
\colhead{} & \colhead{(d)} & \colhead{($R_{\oplus}$)} & \colhead{(Reference)} & \colhead{(This work)} } 

\startdata
1 & 0.4--50 & 1--1.5 & $0.56^{+0.06}_{-0.05}$ & 0.71$\pm$0.09 \\
1 & 0.4--50 & 1.5--2 & $0.46^{+0.07}_{-0.05}$ & 0.53$\pm$0.05 \\
1 & 0.4--200 & 1--4 & 2.5$\pm$0.2 & 1.59$\pm$0.14 \\
2 & 0.4--50 & 0.5--2 & 0.81$\pm$0.19 & 0.65$\pm$0.06 \\
3 & 0.5--10 & 0.5--2.5 & $1.19^{+0.70}_{-0.49}$ & 0.16$\pm$0.02 \\
4 & 0.4--7 & 0.5--2 & $0.61^{+0.24}_{-0.19}$ & 0.12$\pm$0.01 \\
\enddata
\tablecomments{Our calculations included the same period and radius range of the reference, but there are some differences in the planet hosts. This is particularly notable for \citet{Hardegree-Ullman2019} and \citet{Ment2023}, which went deeper into the mid and late-type M dwarf range than our Kepler/K2 survey with relatively small planet sample sizes of 13 and 7, respectively.}

\tablerefs\ (1) \citet{Dressing2015}, (2) \citet{Mulders2015a}, (3) \citet{Hardegree-Ullman2019}, (4) \citet{Ment2023}
\end{deluxetable*}

\subsection{TESS, Roman, and PLATO}\label{sec:tessroman}

To date, there are $\sim$300 TESS Objects of Interest (TOIs) between 1--4~$R_{\oplus}$ orbiting probable M dwarfs down to $T_{\mathrm{eff}}\approx2800$~K.\footnote{Data from ExoFOP, \url{https://exofop.ipac.caltech.edu/tess/view_toi.php}, accessed 3 Jun 2025.} A thorough and uniform analysis of TESS data, on par with the Scaling K2 project, could significantly expand the M dwarf small planet sample through mid-type M dwarfs and further constrain the super-Earth and sub-Neptune trends identified here. 

NASA's upcoming Nancy Grace Roman Space Telescope will also push exoplanet occurrence rate calculations toward mid and late-type M dwarfs. The Roman Galactic Bulge Time Domain Survey (GBTDS) parameters were motivated by the requirements of an exoplanet microlensing study \citep{Penny2019}, but are fortuitously favorable to make the GBTDS a substantial exoplanet transit survey, enabling yields between 60,000 and 200,000 transiting exoplanets. This includes an estimated $\sim$1000--3500 planets smaller than $4 R_{\oplus}$ around M dwarfs \citep{Montet2017,Wilson2023}. While transiting planets smaller than $\sim$2\,$R_{\oplus}$ are not expected to be easily detectable around F and G stars, they will be for M dwarfs and bright K dwarfs (see \citealt{Wilson2023}, Figure 12 and \citealt{Tamburo2023}, Figure 8). Further, the Roman \textit{F}146 bandpass is redder than previous surveys ($\sim$0.9--2~$\mu$m, compared to $\sim$0.6--1~$\mu$m for TESS and $\sim$0.4--0.9~$\mu$m for Kepler) and is ideal to capture the flux from M dwarfs. A single 72-day GBTDS survey season is comparable in length to each K2 campaign. This baseline could be extended and aid in longer-period planet detection if the currently planned total of six observing seasons is executed. 

The upcoming European Space Agency (ESA) PLAnetary Transits and Oscillations (PLATO) mission \citep{Rauer2025} will search for transiting planets around nearby stars, with two long stares of at least two continuous years the northern and southern hemispheres, and covering regions of 2149 deg$^2$ \citep{Nascimbeni2025}. While M dwarfs are not the main focus of PLATO, at least 5000 M dwarfs are expected to be monitored \citep{Montalto2021}. We encourage observing as many M dwarfs as possible with PLATO to enable the broadest range of exoplanet demographic studies with this population. The long baseline of PLATO will extend the period range for which M dwarf planets can be searched. The first proposed long-stare observing field for PLATO overlaps with a majority of the TESS southern continuous viewing zone \citep{Nascimbeni2025}, which could further extend the period baseline for a significant fraction of targets.

\section{Conclusions}\label{sec:conclusions}

Combining data from Kepler and its successor mission K2 affords us the unique opportunity to not only expand our planet sample, but also look for and identify trends not possible with the individual datasets. In this paper we:
\begin{itemize}
    \item Used empirical spectrophotometric calibrations to place the Kepler (94,422) and K2 (90,344) FGKM main sequence stellar populations on the same scale.
    \item Computed the combined Kepler and K2 planet radius valley in both orbital period/radius and incident flux/radius space for FGKM hosts.
    \item Computed exoplanet occurrence rates as a function of stellar host temperature for super-Earths and sub-Neptunes.
    \item Identified a peak in short period ($P=1-40$ days) sub-Neptune occurrence rates at a host star temperature of $3750^{+153}_{-97}$~K, with a rapid drop toward cooler host stars. This peak is qualitatively consistent with several recent planet formation models.
    \item Identified a drop in sub-Neptune occurrence rates beyond $5758^{+94}_{-145}$~K. This drop will be the subject of a future Scaling K2 analysis.
\end{itemize}

These results show the power of homogeneous analysis across surveys. This study, alongside the entire Scaling K2 project, can be used as a template for future combined transit demographics analyses. Additional data from NASA missions like TESS and Roman and ESA's PLATO will delve deeper into regions of stellar and planet parameter space not well sampled by Kepler and K2. Our observational results also provide key constraints to improve future planet formation modeling.

\begin{acknowledgments}
This research has made use of the NASA Exoplanet Archive, which is operated by the California Institute of Technology, under contract with the National Aeronautics and Space Administration under the Exoplanet Exploration Program. This research has made use of the Exoplanet Follow-up Observation Program (ExoFOP; DOI: 10.26134/ExoFOP5) website, which is operated by the California Institute of Technology, under contract with the National Aeronautics and Space Administration under the Exoplanet Exploration Program. This research made use of the cross-match service provided by CDS, Strasbourg.
\end{acknowledgments}

\begin{contribution}

KKH-U derived stellar and exoplanet parameters, computed planet occurrence rates, and wrote a majority of the manuscript. GJB and JKZ developed code for planet occurrence rates and power-law fitting, and provided additional text and analysis for occurrence rates in the manuscript. JLC developed and obtained funding for the initial Scaling K2 project and provided key input and edits to the manuscript. SB, KMB, RBF, SG, and PRK contributed to the science ideas and text presented herein.

\end{contribution}

\facilities{Exoplanet Archive, ExoFOP, Gaia}

\software{\texttt{astropy} \citep{astropy:2013, astropy:2018, AstropyCollaboration2022}, \texttt{ExoMult} \citep{Zink2019}, \texttt{matplotlib} \citep{Hunter:2007}, \texttt{numpy} \citep{harris2020array}, \texttt{pandas} \citep{mckinney-proc-scipy-2010,reback2020pandas}}

\appendix
\section{Stellar and Planet Parameters\label{app:A}}

Here we present the catalogs of Kepler stars (Table~\ref{tab:kep}), K2 stars (Table~\ref{tab:dbpl}), Kepler planet candidates (Table~\ref{tab:kepplanet}), and K2 planet candidates (Table~\ref{tab:k2planet}) that were used in the analysis presented herein. Information about the derivation of the stellar and planet parameters can be found in Sections~\ref{sec:sample} and \ref{sec:planets}, respectively.

\begin{deluxetable*}{clccrccc}[hb!]
\tablecaption{Properties of Kepler Stars\label{tab:kep}}

\tablehead{\colhead{Kepler ID} & \colhead{Distance} & \colhead{$T_{\mathrm{eff}}$} & \colhead{$\log g$} & \colhead{[Fe/H]} & \colhead{$L_{\star}$} & \colhead{$R_{\star}$} & \colhead{$M_{\star}$} \vspace{-0.75em}\\ 
\colhead{} & \colhead{(pc)} & \colhead{(K)} & \colhead{[cm s$^{-2}$]} & \colhead{(dex)} & \colhead{($L_{\odot}$)} & \colhead{($R_{\odot}$)} & \colhead{($M_{\odot}$)}
} 

\startdata
757450 & $796.55_{-10.78}^{+14.24}$ & $4925\pm144$ & $4.414\pm0.079$ & $0.009\pm0.217$ & $0.492\pm0.043$ & $0.960\pm0.066$ & $0.824\pm0.067$ \\
892195 & $452.44_{-2.90}^{+3.22}$ & $5169\pm151$ & $4.421\pm0.069$ & $0.239\pm0.217$ & $0.587\pm0.047$ & $0.958\pm0.062$ & $0.864\pm0.070$ \\
892203 & $558.51_{-3.10}^{+3.68}$ & $5504\pm161$ & $4.381\pm0.064$ & $0.244\pm0.217$ & $0.931\pm0.064$ & $1.063\pm0.069$ & $0.964\pm0.077$ \\
892675 & $574.30_{-3.78}^{+3.67}$ & $5707\pm167$ & $4.375\pm0.079$ & $-0.057\pm0.217$ & $1.120\pm0.094$ & $1.091\pm0.080$ & $1.012\pm0.082$ \\
892718 & $841.54_{-21.79}^{+18.19}$ & $4853\pm142$ & $4.564\pm0.086$ & $-0.028\pm0.217$ & $0.271\pm0.028$ & $0.734\pm0.058$ & $0.716\pm0.059$ \\
892832 & $778.76_{-12.78}^{+21.03}$ & $4884\pm143$ & $4.552\pm0.075$ & $-0.101\pm0.217$ & $0.295\pm0.026$ & $0.765\pm0.056$ & $0.733\pm0.060$ \\
892834 & $562.23_{-6.77}^{+6.33}$ & $4724\pm138$ & $4.557\pm0.068$ & $-0.085\pm0.217$ & $0.237\pm0.018$ & $0.734\pm0.050$ & $0.695\pm0.056$ \\
892882 & $601.10_{-7.00}^{+5.86}$ & $4983\pm146$ & $4.516\pm0.067$ & $-0.086\pm0.217$ & $0.321\pm0.027$ & $0.769\pm0.055$ & $0.745\pm0.061$ \\
892911 & $1580.49_{-49.34}^{+44.91}$ & $5403\pm158$ & $4.237\pm0.070$ & $-0.114\pm0.217$ & $1.161\pm0.134$ & $1.241\pm0.097$ & $1.016\pm0.084$ \\
893033 & $516.58_{-4.05}^{+4.41}$ & $4703\pm138$ & $4.700\pm0.277$ & $-0.359\pm0.217$ & $0.185\pm0.015$ & $0.653\pm0.048$ & $0.681\pm0.356$ \\
\enddata

\tablecomments{This table is available in its entirety in machine-readable form.}
\end{deluxetable*}

\begin{deluxetable*}{cclcrcccc}
\tablecaption{Properties of K2 Stars\label{tab:k2}}

\tablehead{\colhead{EPIC ID} & \colhead{Distance} & \colhead{$T_{\mathrm{eff}}$} & \colhead{$\log g$} & \colhead{[Fe/H]} & \colhead{$L_{\star}$} & \colhead{$R_{\star}$} & \colhead{$M_{\star}$}\vspace{-0.75em} \\ 
\colhead{} & \colhead{(pc)} & \colhead{(K)} & \colhead{[cm s$^{-2}$]} & \colhead{(dex)} & \colhead{($L_{\odot}$)} & \colhead{($R_{\odot}$)} & \colhead{($M_{\odot}$)}
} 

\startdata
201123619 & $200.86_{-1.91}^{+1.72}$ & $3729\pm109$ & $4.758\pm0.031$ & $-0.633\pm0.217$ & $0.048\pm0.002$ & $0.500\pm0.015$ & $0.519\pm0.019$ \\
201124275 & $532.70_{-5.77}^{+7.51}$ & $5942\pm174$ & $4.706\pm0.079$ & $-1.779\pm0.217$ & $0.507\pm0.047$ & $0.674\pm0.052$ & $0.833\pm0.068$ \\
201126351 & $288.96_{-2.37}^{+2.60}$ & $4323\pm126$ & $4.857\pm0.282$ & $-0.370\pm0.217$ & $0.155\pm0.009$ & $0.702\pm0.047$ & $1.227\pm0.744$ \\
201126368 & $246.80_{-1.46}^{+1.64}$ & $4038\pm118$ & $4.626\pm0.026$ & $-0.229\pm0.217$ & $0.095\pm0.004$ & $0.634\pm0.018$ & $0.625\pm0.020$ \\
201127502 & $248.31_{-1.37}^{+1.48}$ & $6162\pm181$ & $4.328\pm0.088$ & $-0.113\pm0.217$ & $1.943\pm0.166$ & $1.228\pm0.110$ & $1.152\pm0.094$ \\
201128834 & $290.26_{-3.31}^{+2.98}$ & $3881\pm113$ & $4.761\pm0.029$ & $-0.896\pm0.217$ & $0.054\pm0.003$ & $0.508\pm0.016$ & $0.532\pm0.021$ \\
201129780 & $337.92_{-8.68}^{+7.99}$ & $3378\pm99$ & $4.753\pm0.035$ & $0.155\pm0.217$ & $0.032\pm0.002$ & $0.480\pm0.017$ & $0.477\pm0.018$ \\
201129879 & $455.60_{-5.07}^{+5.16}$ & $4894\pm143$ & $4.577\pm0.072$ & $-0.025\pm0.217$ & $0.260\pm0.020$ & $0.719\pm0.051$ & $0.709\pm0.058$ \\
201129911 & $501.57_{-7.36}^{+6.74}$ & $4978\pm146$ & $4.556\pm0.069$ & $-0.091\pm0.217$ & $0.318\pm0.024$ & $0.754\pm0.052$ & $0.745\pm0.060$ \\
201129933 & $435.45_{-5.71}^{+4.42}$ & $4901\pm143$ & $4.523\pm0.067$ & $0.053\pm0.217$ & $0.334\pm0.028$ & $0.803\pm0.055$ & $0.759\pm0.062$ \\
\enddata

\tablecomments{This table is available in its entirety in machine-readable form.}
\end{deluxetable*}

\begin{deluxetable*}{cccrc}
\tablecaption{Kepler Super-Earth and Sub-Neptune Exoplanet Parameters\label{tab:kepplanet}}

\tablehead{\colhead{Kepler ID} & \colhead{KOI} & \colhead{Radius} & \colhead{Period} & \colhead{Incident Flux} \vspace{-0.75em}\\ 
\colhead{} & \colhead{} & \colhead{($R_\Earth$)} & \colhead{(days)} & \colhead{($S_\Earth$)} } 

\startdata
1026957 & 0958.01 & $2.457\pm0.187$ & $21.621\pm0.776$ & $13.445\pm1.297$ \\
1717722 & 3145.01 & $1.605\pm0.280$ & $4.561\pm0.131$ & $73.281\pm41.229$ \\
1718189 & 0993.01 & $1.661\pm0.275$ & $21.854\pm0.588$ & $28.611\pm3.078$ \\
1718189 & 0993.02 & $1.470\pm0.350$ & $12.962\pm0.318$ & $55.709\pm5.989$ \\
1721157 & 4644.01 & $1.088\pm0.126$ & $3.439\pm0.101$ & $163.134\pm15.762$ \\
1724719 & 4212.02 & $1.059\pm0.197$ & $9.997\pm0.303$ & $185.554\pm22.942$ \\
1871483 & 6256.01 & $1.594\pm0.202$ & $14.093\pm0.509$ & $25.858\pm2.723$ \\
1872821 & 2351.01 & $1.651\pm0.152$ & $10.268\pm0.298$ & $72.396\pm7.807$ \\
2142522 & 2403.01 & $1.316\pm0.121$ & $13.327\pm0.278$ & $133.521\pm12.153$ \\
2161536 & 2130.01 & $1.844\pm0.130$ & $16.923\pm0.322$ & $7.967\pm0.538$ \\
\enddata

\tablecomments{This table is available in its entirety in machine-readable form.}

\end{deluxetable*}

\begin{deluxetable*}{cllccccc}
\tablecaption{K2 Super-Earth and Sub-Neptune Exoplanet Parameters\label{tab:k2planet}}
\tablehead{\colhead{EPIC ID} & \colhead{Candidate} & \colhead{Radius} & \colhead{Period} & \colhead{Incident Flux} \vspace{-0.75em}\\ 
\colhead{} & \colhead{} & \colhead{($R_\Earth$)} & \colhead{(days)} & \colhead{($S_\Earth$)} } 

\startdata
201155177 & 201155177.01 & $2.216\pm0.281$ & $6.689\pm0.0008$ & $54.049\pm5.849$ \\
201208431 & 201208431.01 & $2.468\pm0.247$ & $10.0042\pm0.0009$ & $17.878\pm0.899$ \\
201338508 & 201338508.01 & $2.002\pm0.214$ & $10.9349\pm0.0011$ & $12.456\pm0.684$ \\
201357643 & 201357643.01 & $3.943\pm0.292$ & $11.8911\pm0.0003$ & $140.066\pm15.106$ \\
201384232 & 201384232.01 & $2.336\pm0.221$ & $30.9434\pm0.0034$ & $21.448\pm2.459$ \\
201390927 & 201390927.01 & $3.566\pm0.699$ & $2.6378\pm0.0002$ & $194.341\pm16.458$ \\
201403446 & 201403446.01 & $2.471\pm0.251$ & $19.1533\pm0.0029$ & $115.17\pm11.57$ \\
201445392 & 201445392.01 & $2.347\pm0.308$ & $5.0646\pm0.0004$ & $91.851\pm9.704$ \\
201445392 & 201445392.02 & $2.799\pm0.342$ & $10.3535\pm0.0007$ & $34.441\pm3.902$ \\
201456770 & 201456770.01 & $2.321\pm0.299$ & $8.0069\pm0.0013$ & $98.416\pm9.982$ \\
\enddata

\tablecomments{This table is available in its entirety in machine-readable form.}
\end{deluxetable*}

\clearpage
\bibliography{main}{}
\bibliographystyle{aasjournal}

\end{document}